\documentclass{article}
\usepackage{authblk}
\usepackage{graphicx}
\usepackage{spconf,amsmath,graphicx}
\usepackage{amssymb}
\usepackage{xcolor}
\usepackage{booktabs}
\usepackage{array}
\usepackage{booktabs}
\usepackage{array}
\usepackage{subcaption}
\newcolumntype{?}{!{\vrule width 0.08em}}
\usepackage[pagebackref=true,breaklinks=true,letterpaper=true,colorlinks,bookmarks=false,citecolor=blue,linkcolor=blue]{hyperref}

\title{Self-supervised Audio Spatialization with correspondence Classifier}
\name{Yu-Ding Lu, Hsin-Ying Lee, Hung-Yu Tseng, Ming-Hsuan Yang}
\address{University of California at Merced} 
\begin{document}
%
\maketitle
\begin{abstract}
Spatial audio is an essential medium to audiences for 3D visual and auditory experience.
However, the recording devices and techniques are expensive or inaccessible to the general public.
In this work, we propose a self-supervised audio spatialization network that can generate spatial audio given the corresponding video and monaural audio.
To enhance spatialization performance, we use an auxiliary classifier to classify ground-truth videos and those with audio where the left and right channels are swapped. 
We collect a large-scale video dataset with spatial audio to validate the proposed method.
Experimental results demonstrate the effectiveness of the proposed model on the audio spatialization task.

\end{abstract}
\begin{keywords}
Audio-visual, Spatial audio, \\ 
Self-supervised
\end{keywords}

%

%
\vspace{-3mm}
\section{Introduction}
\label{sec:intro}
\vspace{-2mm}

Humans perceive the environments through various different sensory systems.
Among all senses, visual and auditory cues are particularly important since they both provide \textit{spatial} information.
Mankind can effortlessly learn from observation the correlation and co-occurrence between vision and audio.
For example, upon seeing a car racing from the right side, we will expect to hear the engine sound from the right to the left.
%
%

Recently, there has been a growing interest in exploiting cross-modality information between visual and auditory domains.
One line of work targets at self-supervised learning, which aims to learn feature representations by solving surrogate tasks defined from the structure of raw data~\cite{doersch2015unsupervised,lee2017unsupervised,noroozi2016unsupervised}. 
The correlations between images and sounds serve as natural supervisory signals for representation learning~\cite{aytar2016soundnet,owens2016ambient,aytar2017seehearread,owens2018audio}.
The other line of work makes use of the spatial property of vision and audio to localize the source of sounds in images~\cite{arandjelovic2017objects, senocak2018learning}, or separate different audio-visual sources~\cite{ephrat2018looking,gao2018learning,owens2018audio}.
In this work, we focus on audio spatialization, a technique to generate spatial audio from non-spatial audio, considering the spatial and temporal correlation between videos and audios.



Spatial audio, also known as binaural audio, is recorded by simulating the way surrounding sounds transmits to human ears.
Spatial audio provides listeners a sense of space beyond conventional stereo audio and allows users to precisely pinpoint the direction of the source of sounds.
However, the cost of binaural recording is expensive and recording technique is non-trivial. 
It is not easy for the general public to record spatial audio by themselves. 
Additionally, most video recording devices are with single microphones, which records monaural (mono) audio. 
Thus, the audio-visual experience is limited.

%
%


%
%
Recently, Pedro et al.~\cite{morgado2018self} propose a self-supervised neural network architecture to separate sound sources from a mixed audio input and $360^{\circ}$ videos. Specifically, this model can generate spatial audio which enables users to experience sound in all directions.
%
This network architecture has been extended  to  upconvert a single mono recording into spatial audio guided by encoding both appearance (RGB frames) and motion (optical flow \cite{ilg2017flownet}). 
Specifically, the network converts mono audio into first-order ambisonics, which consists of four channels that store the first-order coefficients to simulate $360^{\circ}$ surrounding sound.
However, most videos and audios are not recorded in the $360^{\circ}$ format.
The method proposed in~\cite{morgado2018self} cannot properly apply to general videos.
%

In this work, we propose an audio spatialization network (ASN), a self-supervised framework for audio spatialization.
Specifically, the proposed method aims to generate binaural audio given the corresponding video and mono audio.
Instead of predicting raw waveform, the proposed model learns to map from mono audio to Ideal Ratio Mask (IRM), which has been used in speech enhancement with stable and effective performance~\cite{narayanan2013ideal, sun2018novel}, in the spatial time-frequency (T-F) domain.
%
%
%
%
%
%
Furthermore, inspired by ACGAN~\cite{odena2017conditional}, which leverages an additional correspondence classifier to help the generation task, we apply a binary classifier to distinguish videos with correct audios and videos with audio where the left and right channels are swapped. 
%
The classifier provides auxiliary training signal to improve the performance.

For this audio spatialization task, we collect videos in the wild totaling 10 hours.
The dataset contains two categories: racing and music.
For evaluation, we compare with the baseline methods on short-time Fourier transform (STFT) distance and envelope distance (ENV).
The quantitative results demonstrate the effectiveness of the proposed method.

\begin{figure*}[t]

\centering
\renewcommand{\tabcolsep}{1pt} 
\renewcommand{\arraystretch}{1} 
\begin{center}
\begin{tabular}{c}
 \includegraphics[width=0.9\linewidth]{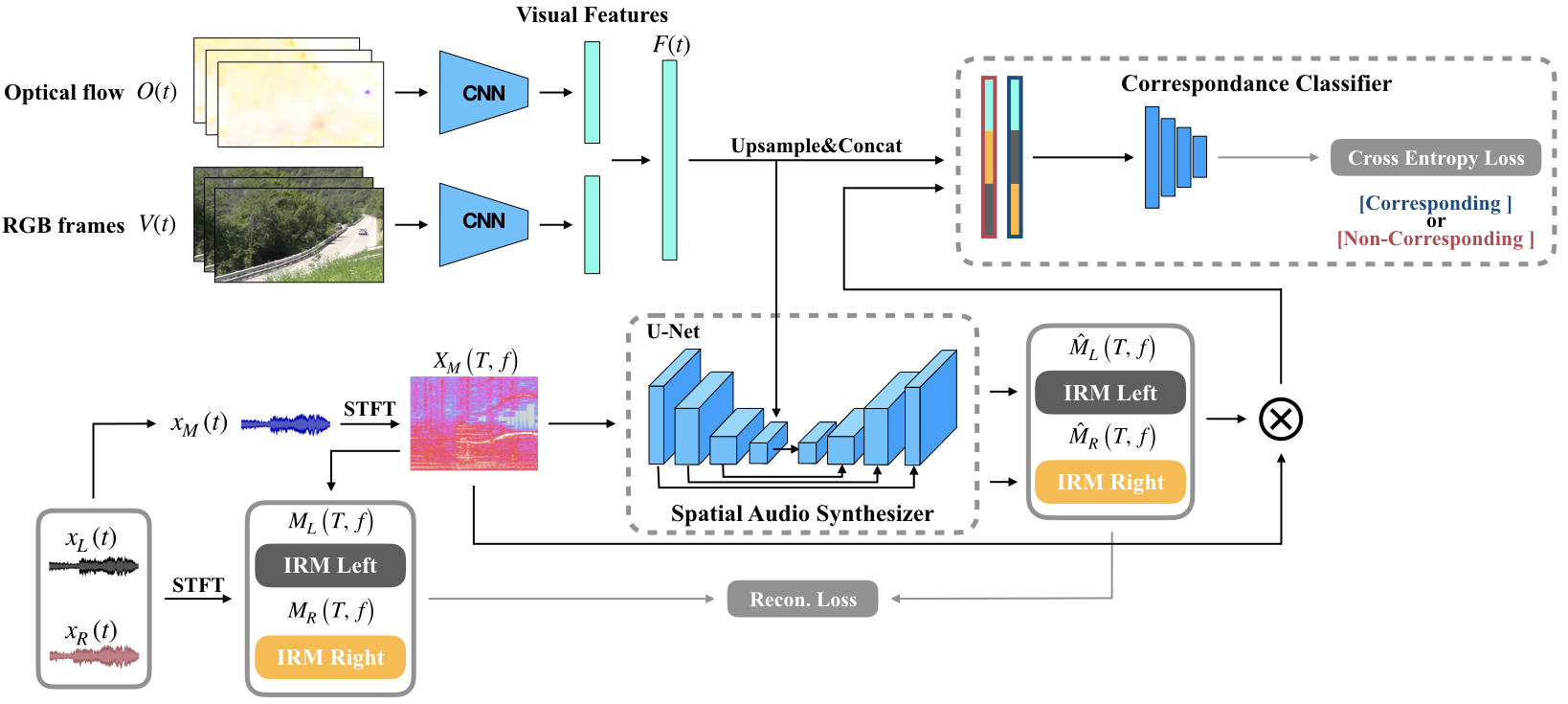} \\
\end{tabular}
\end{center}
\vspace{-6mm}
\caption{\textbf{Network architecture.} 
The proposed model takes audio-visual features $F(t)$ (concatenated by optical flow features $O(t)$ and RGB features $V(t)$) to convert mono audio $x_M(t)$ into left and right ideal ratio mask  $\hat{M}_L$ and $\hat{M}_R$. 
The predicted ratio masks are then used to reconstruct spatial audio.
The model consists of two modules: the spatial audio synthesizer (Section \ref{subsec:network}) predicting ratio masks and the correspondence classifier (Section \ref{subsec:classifier}) providing auxiliary training signal.
}

\label{fig:model}
\vspace{-4mm}
\end{figure*}

The contributions of this work are summarized as follows.
First, we proposed an audio spatialization network with auxiliary classifier. The proposed model can generate spatial audio given the corresponding video and mono audio.
Second, we collect a large-scale video dataset with spatial audio, which will be released to the public along with our code, upon the acceptance of this paper. 
%

\section{PROPOSED METHOD}
\label{sec:method}
Our goal is to generate spatial audio given original video and mono audio.
As shown in Fig. \ref{fig:model}, the model consists of a spatial audio synthesizer network and a binary correspondence classifier.%
The synthesizer network takes as inputs the optical flow and RGB features as well as the audio features to generate left and right IRMs.
The binary correspondence classifier distinguishes videos with correct left and right IRM and videos with swapped IRMs. 
This auxiliary classifier helps the synthesizer to generate realistic and correct masks.

In this section, we first discuss the feature extraction process, and then introduce the synthesizer network. Finally, we detail the auxiliary classifier. 
%

\vspace{-2mm}
\subsection{Audio and visual feature extraction}
\label{subsec:feature}
\textbf{Audio features.} Audio features are extracted by STFT (short-time Fourier transform) to the T-F domain. In general, audio representation in T-F domain is successfully used for many tasks such as audio classification \cite{hershey2017cnn} and speech enhancement \cite{lu2013speech, fu2017complex}. We thus extract spectrogram as audio features. First, STFT is applied on input audio and spatial audio to obtain the spectrograms. Then the magnitude of spectrograms are transformed into log-frequency scale to serve as the inputs to our model.

\noindent\textbf{Video features.}
Similar to \cite{morgado2018self}, we extract video features by using a two-stream of ResNet-18 network. 
The two-stream ResNet-18 network extract features of RGB frames and optical flow (predicted by FlowNet~\cite{IMKDB17, lai2018learning}).
Both streams are pretrained by ImageNet for classification. 
We then flatten and concatenate both features.

\noindent\textbf{Audio-Visual features.}
Audio features are extracted at a higher frame rate than video features. To synchronize audio features and video features, we use 44 fps for audio features and upsample video features from 10 fps to 44 fps by the nearest neighbor algorithm. 
We then concatenate video features with the encoded audio feature as inputs to the up-sampling part of the synthesizer network.


\vspace{-2mm}
\subsection{Spatial audio synthesizer network}
\label{subsec:network}
The goal of audio spatialization is to generate binaural channels $x_L(t), x_R(t)$ from monaural audio $x_M(t)$, optical flow $O(t)$ and synchronized video $V(t)$. 
$O(t)$ and $V(t)$ are encoded and concatenated to visual features $F(t)$.

Spatial signal consists of two channels recorded by simulating how human perceive stereo sound.
We mix spatial channels into mono channels as a self-supervisory training signal for the synthesizer network.
Let $x_L(t)$ and $x_R(t)$ be waveform of the left and right channels of spatial audio and $V(t)$ be their corresponding video. 
We then mixed two channels into single channel $x_m(t) = x_L(t) + x_R(t)$. 
With STFT, we get $X_L(T, f)$, $X_R(T, f)$, and $X_M(T, f)$, the energy signals in the $T$ time frame and the $k$ frequency bin of the left, right, and the mixed channel, respectively. 
Then, the ideal ratio masks for left and right of spatial audio are defined as
\begin{equation}
M_{L}\left( T,\; f \right)\; =\; \sqrt{\frac{X_{L}\left( T,\; f \right)^{2}}{X_{R}\left( T,\; f \right)^{2}\; +\; X_{L}\left( T,\; f \right)^{2}}}
\end{equation}

\begin{equation}
M_{R}\left( T,\; f \right)\; =\; \sqrt{\frac{X_{R}\left( T,\; f \right)^{2}}{X_{R}\left( T,\; f \right)^{2}\; +\; X_{L}\left( T,\; f \right)^{2}}}
\end{equation}

The synthesizer $S$ is designed as a U-net~\cite{ronneberger2015u} architecture, which has been widely applied to audio tasks \cite{jansson2017singing}.
The synthesizer takes $X_M(T, f)$ and $F(t)$ as inputs and reconstruct left and right IRM: $ \{\hat{M_L}, \hat{M_R}\} = S(X_M, F)$.
The reconstruction loss:
\begin{equation}
    l_{\mathrm{recon}} = \left\lVert{M_L - \hat{M_L}}\right\rVert_{1}+\left\lVert{M_R - \hat{M_R}}\right\rVert_{1}
\end{equation}


\vspace{-2mm}
\subsection{Correspondence classifier}
\label{subsec:classifier}
To help synthesizer better learn the correspondences between audio and visual features, we apply an auxiliary correspondence classifier.
The classifier is trained to distinguish between videos with correct corresponding left and right signals $\{F(t), {X_L}, {X_R}\}$ and videos with swapped signals $\{F(t), {X_R}, {X_L}\}$.
The classifier is then used to help the training of the synthesizer.
We first reconstruct the energy signal on the T-F domain:
\begin{equation}
\begin{aligned}
	&\hat{X_L}\left( T,\; f \right) = \hat{M_L}\left( T,\; f \right) \times X_m \left( T,\; f \right)\\
	&\hat{X_R}\left( T,\; f \right) = \hat{M_R}\left( T,\; f \right) \times X_m \left( T,\; f \right)
	\end{aligned}
\end{equation}
then the classifier is used to distinguish between videos with correct corresponding left and right signals $\{F(t), \hat{X_L}, \hat{X_R}\}$ and videos with swapped signals $\{F(t), \hat{X_R}, \hat{X_L}\}$.
The classifier provides training signal with the cross-entropy loss $L_{\mathrm{cls}}$.

The full objective of training the synthesizer is:
\begin{equation}
    L = L_{\mathrm{recon}} + \lambda_{cls}L_{\mathrm{cls}}
\end{equation}
where $\lambda_{cls}L_{\mathrm{cls}}$ is the weight to control the importance of the classification term.
\vspace{-2mm}
\section{Experimental Results}
\label{sec:experiment}

\vspace{-2mm}
\subsection{Datasets}
In order to train our model, we collect a video dataset consisting of two categories of video with spatial audio including different sound sources, including playing music and racing car. These videos are collected in-the-wild from YouTube using keywords related to spatial audio, e.g., spatial audio, ambisonic, binaural and 3D audio. Two categories are denoted SP-RACING and SP-MUSIC. The details are list in Table \ref{table:dataset}.

\begin{table}[t]
    \caption{
       \textbf{Statistics of datasets.} The total length of the two categories of the collected dataset.
    }
\centering
\begin{tabular}{c c c }

\toprule
Dataset & \textbf{SP-RACING} & \textbf{SP-MUSIC}  \\ \midrule
Time    & 6.8 hrs         & 2.5 hrs       \\ \bottomrule
\end{tabular}
\vspace{-3mm}
\label{table:dataset}
\end{table}

\begin{figure}[thb]
\includegraphics[width=\linewidth]{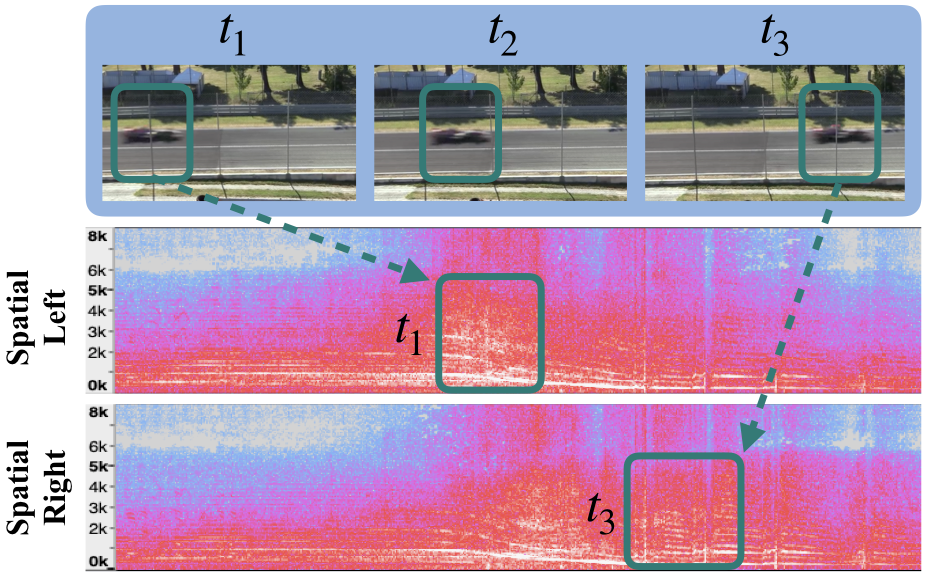}
\caption{\textbf{Examples of the collected dataset.}
We show an example SP-RACING video clip and its corresponding left and right time-frequency energy signals.
The racing car moving from the left to the right (from $t1$ to $t3$) is reflected on the energy signals in T-F domain. 
}
\label{fig:dataset}
\vspace{-3mm}
\end{figure}

\subsection{Implementation details}
Our model is implemented in PyTorch on a NVIDIA GeForce GTX 1080ti GPU.
For our experimental settings, we randomly choose 90\% of videos for training and 10\% for testing.
The audio is at 44.1kHz and video at 10fps. 
The STFT is computed by using FFT size of 2048, a Hann window of 40ms and hop length of 10ms. 
The RGB and flow features are both 1025 dimensions.
Each training sample consists of a chunk of about 1 sec of mono audio and video features (RGB and mono). 
For data augmentation, we randomly flip video and exchange the left and right channels of spatial audio at the same time.
We use Adam optimizer with parameters $\beta_1$ = 0.9, $\beta_2$ = 0.999, weight decay = $1e$$-4$, learning rate of $5e$$-5$ and batch size of 44. 
In all experiment, we set the hyper-parameter $\lambda_{cls}=0.1$. 

As for network architecture of the synthesizer, we apply the U-Net to encode the mono spectrogram, which reduces the spectrogram dimensionality and extracts high-level features. 
U-Net consists of a number of convolutional downsampling layers distilling higher level features and several upsampling layers where skip connections are added. 
We follow \cite{morgado2018self} to set all layers of our U-Net synthesizer. After the layer of this U-Net, we append two Sigmoid layer to predict spatial ratio masks.
As for the classifier, it consists of three fully-connected layer with ReLU and one Sigmoid layer. 

\begin{figure*}[t]
\centering
\renewcommand{\tabcolsep}{1pt} 
\renewcommand{\arraystretch}{1} 
\begin{center}
\begin{tabular}{c}
 \includegraphics[width=0.95\linewidth]{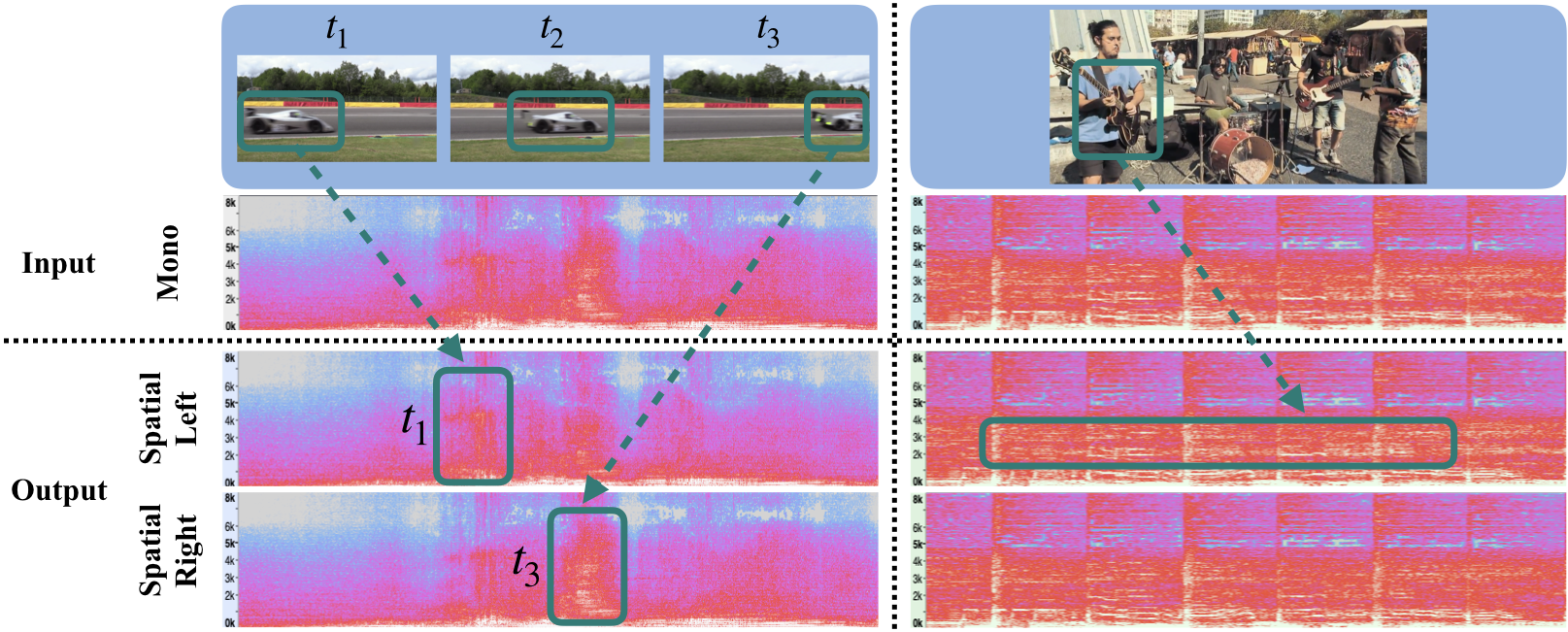} \\
\end{tabular}
\end{center}
\vspace{-3mm}
\caption{\textbf{Results predicted by the proposed audio spatialization network.}
We present predicted results from the SP-RACING (left) and the SP-MUSIC (right).
(Left) The energy signal in T-F domain reflects the racing car moving from the left to the right.
(Right) The energy signal in T-F domain reflects the musician on the left, who is the dominant source of sounds.
} 
\label{fig:result}
\vspace{-3mm}
\end{figure*}

\begin{table}[thb]
    \centering
    
    \caption{
       \textbf{Quantitative results.}
       We evaluate ASN on two settings:
       (a) On all testing videos, and
       (b) On testing clips where the ratio of left and right channels is larger than a threshold.
    }
    \label{tab:result}

    \begin{tabular}{ccccc}
        & \multicolumn{2}{c}{SP-MUSIC}
        & \multicolumn{2}{c}{SP-RACING}\\
        \cline{2-5}
        & STFT & ENV & STFT & ENV \\
        \midrule[0.08em]
        MONO  & 0.963 & 0.407 & 0.864 & 0.218 \\
        ASN w/o classifier  & 0.898 & 0.375 & 0.797 & 0.202\\
        ASN (ours)  & \textbf{0.850} & \textbf{0.351} & \textbf{0.744} & \textbf{0.192}\\
        \bottomrule[0.08em]
    \end{tabular}
    \subcaption{Results on all testing videos}

     \begin{tabular}{ccccc}
        & \multicolumn{2}{c}{SP-MUSIC}
        & \multicolumn{2}{c}{SP-RACING}\\
        \cline{2-5}
        & STFT & ENV & STFT & ENV \\
        \midrule[0.08em]
        MONO  & 1.329 & 0.812 & 1.397 & 0.627 \\
        ASN w/o classifier  & 1.237 & 0.755 & 1.313 & 0.594\\
        ASN (ours)  & \textbf{1.133} & \textbf{0.664} & \textbf{1.223} & \textbf{0.556}\\
        \bottomrule[0.08em]
    \end{tabular}
    \subcaption{Results on filtered testing clips}
\end{table}

\subsection{Evaluation metrics}
In this paper, we evaluate our results on time-domain and frequency-domain of audio signals following \cite{morgado2018self}.

\noindent
\textbf{STFT distance}:
This metric computes the Euclidean distance between the ground-truth and predicted spectrograms of left and right channels in the T-F domain. 
\begin{equation}
D_{\textrm{STFT}}\; =\; \sum_{p\; \in \; \left\{ L,\; R \right\}}^{}{\left| \left| X_{p}\; -\; \tilde{X_{p}} \right| \right|^{2}}
\end{equation}
where $\| \cdot \|$ is the Euclidean norm.

\noindent
\textbf{Envelope (ENV) distance}:
In time-domain, we can measure the Euclidean distance between the envelopes of real spatial audio and our results. Instead of STFT distance, ENV distance is able to capture perceptual similarity.
\begin{equation}
D_{\textrm{ENV}}\; =\; \sum_{p\; \in \; \left\{ L,\; R \right\}}^{}{\left| \left| ENV[x_{p}]\; -\; ENV[\tilde{x_{p}}] \right| \right|^{2}}
\end{equation}
where $x$ is raw waveform of ground-truth signals and $\tilde{x}$ is predicted waveform of spatial audio. $ENV[x]$ is the envelop of signal $x$.

\subsection{Results and analysis}
We compare the proposed audio spatialization network with two baseline methods.
First, MONO is computed with the mono audio obtained by the input left and right spatial audio.
Second, ASN w/o classifier is to ablate the correspondence classifier.
We evaluate all methods on two settings: evaluation on all testing videos and evaluation on testing clips that are chosen based on the ratio of the left and right channels.
The results are shown in Table \ref{tab:result}.
The synthesizer can generate reasonable spatial audio, while the correspondence classifier can further improve the performance.
In the general setting, ASN outperforms MONO by $12.8\%$ with the STFT and ENV metrics, and outperforms ASN w/o classifier by $6.0\%$ and $5.6\%$ with the STFT and ENV metrics, respectively.
In the filtered setting, ASN outperforms MONO by $13.4\%$ and $14.8\%$, and outperform ASN w/o classifier by $7.6\%$ and $9.2\%$ with the STFT and ENV metrics, respectively.

We demonstrate the qualitative results in Figure \ref{fig:result}.
The energy signals generated by the predicted left and right ideal ratio masks can reflect the movement of the racing cars (left) and the dominant source of sound (right). 


\section{CONCLUSION}
\label{sec:conclusion}
\vspace{-2mm}
In this paper, we propose an audio spatialization network to predict spatial audios from mono audios and the corresponding videos.
The audio spatialzation network consists of a spatial audio synthesizer, which predicts the left and right ideal ratio mask given visual and audio features, and a correspondence classifier, which provide auxiliary training signal to improve the performance.
To validate the effectiveness of the proposed method, we collect a large-scale dataset of videos recorded with spatial audio.
The quantitative and qualitative results show that the proposed framework can generate spatial audios aligned with the video content.
\bibliographystyle{IEEEbib}
\bibliography{strings,refs}

\begin{thebibliography}{10}

\bibitem{doersch2015unsupervised}
Carl Doersch, Abhinav Gupta, and Alexei~A Efros,
\newblock ``Unsupervised visual representation learning by context
  prediction,''
\newblock in {\em ICCV}, 2015.

\bibitem{lee2017unsupervised}
Hsin-Ying Lee, Jia-Bin Huang, Maneesh Singh, and Ming-Hsuan Yang,
\newblock ``Unsupervised representation learning by sorting sequences,''
\newblock in {\em ICCV}, 2017.

\bibitem{noroozi2016unsupervised}
Mehdi Noroozi and Paolo Favaro,
\newblock ``Unsupervised learning of visual representations by solving jigsaw
  puzzles,''
\newblock in {\em ECCV}, 2016.

\bibitem{aytar2016soundnet}
Yusuf Aytar, Carl Vondrick, and Antonio Torralba,
\newblock ``Soundnet: Learning sound representations from unlabeled video,''
\newblock in {\em NIPS}, 2016.

\bibitem{owens2016ambient}
Andrew Owens, Jiajun Wu, Josh~H McDermott, William~T Freeman, and Antonio
  Torralba,
\newblock ``Ambient sound provides supervision for visual learning,''
\newblock in {\em ECCV}, 2016.

\bibitem{aytar2017seehearread}
Yusuf Aytar, Carl Vondrick, and Antonio Torralba,
\newblock ``See, hear, and read: Deep aligned representations,''
\newblock {\em CoRR}, vol. abs/1706.00932, 2017.

\bibitem{owens2018audio}
Andrew Owens and Alexei~A Efros,
\newblock ``Audio-visual scene analysis with self-supervised multisensory
  features,''
\newblock in {\em ECCV}, 2018.

\bibitem{arandjelovic2017objects}
Relja Arandjelovi{\'c} and Andrew Zisserman,
\newblock ``Objects that sound,''
\newblock in {\em ECCV}, 2018.

\bibitem{senocak2018learning}
Arda Senocak, Tae-Hyun Oh, Junsik Kim, Ming-Hsuan Yang, and In~So Kweon,
\newblock ``Learning to localize sound source in visual scenes,''
\newblock in {\em CVPR}, 2018.

\bibitem{ephrat2018looking}
Ariel Ephrat, Inbar Mosseri, Oran Lang, Tali Dekel, Kevin Wilson, Avinatan
  Hassidim, William~T Freeman, and Michael Rubinstein,
\newblock ``Looking to listen at the cocktail party: A speaker-independent
  audio-visual model for speech separation,''
\newblock in {\em SIGGRAPH}, 2018.

\bibitem{gao2018learning}
Ruohan Gao, Rogerio Feris, and Kristen Grauman,
\newblock ``Learning to separate object sounds by watching unlabeled video,''
\newblock in {\em ECCV}, 2018.

\bibitem{morgado2018self}
Pedro Morgado, Nuno Nvasconcelos, Timothy Langlois, and Oliver Wang,
\newblock ``Self-supervised generation of spatial audio for 360° video,''
\newblock in {\em NIPS}, 2018.

\bibitem{ilg2017flownet}
Eddy Ilg, Nikolaus Mayer, Tonmoy Saikia, Margret Keuper, Alexey Dosovitskiy,
  and Thomas Brox,
\newblock ``Flownet 2.0: Evolution of optical flow estimation with deep
  networks,''
\newblock in {\em CVPR}, 2017.

\bibitem{narayanan2013ideal}
Arun Narayanan and DeLiang Wang,
\newblock ``Ideal ratio mask estimation using deep neural networks for robust
  speech recognition,''
\newblock in {\em ICASSP}, 2013.

\bibitem{sun2018novel}
Lei Sun, Jun Du, Tian Gao, Yu-Ding Lu, Yu~Tsao, Chin-Hui Lee, and Neville
  Ryant,
\newblock ``A novel lstm-based speech preprocessor for speaker diarization in
  realistic mismatch conditions,''
\newblock in {\em ICASSP}, 2018.

\bibitem{odena2017conditional}
Augustus Odena, Christopher Olah, and Jonathon Shlens,
\newblock ``Conditional image synthesis with auxiliary classifier gans,''
\newblock in {\em ICML}, 2017.

\bibitem{hershey2017cnn}
Shawn Hershey, Sourish Chaudhuri, Daniel~PW Ellis, Jort~F Gemmeke, Aren Jansen,
  R~Channing Moore, Manoj Plakal, Devin Platt, Rif~A Saurous, Bryan Seybold,
  et~al.,
\newblock ``Cnn architectures for large-scale audio classification,''
\newblock in {\em ICASSP}, 2017.

\bibitem{lu2013speech}
Xugang Lu, Yu~Tsao, Shigeki Matsuda, and Chiori Hori,
\newblock ``Speech enhancement based on deep denoising autoencoder.,''
\newblock in {\em INTERSPEECH}, 2013.

\bibitem{fu2017complex}
Szu-Wei Fu, Ting-yao Hu, Yu~Tsao, and Xugang Lu,
\newblock ``Complex spectrogram enhancement by convolutional neural network
  with multi-metrics learning,''
\newblock in {\em MLSP}, 2017.

\bibitem{IMKDB17}
E.~Ilg, N.~Mayer, T.~Saikia, M.~Keuper, A.~Dosovitskiy, and T.~Brox,
\newblock ``Flownet 2.0: Evolution of optical flow estimation with deep
  networks,''
\newblock in {\em CVPR}, 2017.

\bibitem{lai2018learning}
Wei-Sheng Lai, Jia-Bin Huang, Oliver Wang, Eli Shechtman, Ersin Yumer, and
  Ming-Hsuan Yang,
\newblock ``Learning blind video temporal consistency,''
\newblock in {\em ECCV}, 2018.

\bibitem{ronneberger2015u}
Olaf Ronneberger, Philipp Fischer, and Thomas Brox,
\newblock ``U-net: Convolutional networks for biomedical image segmentation,''
\newblock in {\em MICCAI}, 2015.

\bibitem{jansson2017singing}
Andreas Jansson, Eric Humphrey, Nicola Montecchio, Rachel Bittner, Aparna
  Kumar, and Tillman Weyde,
\newblock ``Singing voice separation with deep u-net convolutional networks,''
\newblock in {\em ISMIR}, 2017.

\end{thebibliography}

\end{document}